\documentclass[english,twocolumn,aps,pra,showpacs]{revtex4}

\usepackage{amssymb}
\usepackage{amsmath}
\usepackage{colordvi}
\usepackage{amsthm}
\usepackage{epsfig}
\usepackage{graphicx}
\usepackage{subfigure}
\usepackage{color}
\usepackage{verbatim}

\usepackage{ulem}

\def\avg(#1){\langle#1\rangle}

\def\be{\begin{equation}}
\def\ee{\end{equation}}
\def\bea{\begin{eqnarray}}
\def\eea{\end{eqnarray}}

\begin{document}

\author{Xi-Wang Luo}
\author{Yu-Na Zhang}
\author{Xingxiang Zhou}
\thanks{email: xizhou@ustc.edu.cn}
\author{Guang-Can Guo}
\author{Zheng-Wei Zhou}
\thanks{email: zwzhou@ustc.edu.cn}
\affiliation{Key Laboratory of Quantum Information, University of Science and Technology
of China, Hefei, Anhui 230026, China}
\affiliation{
Synergetic Innovation Center of Quantum Information and Quantum Physics,
University of Science and Technology of China, Hefei, Anhui 230026, China }

\title{Dynamic phase transitions of a driven Ising chain in a
dissipative cavity}

\begin{abstract}
We study the nonequilibrium quantum phase transition of an Ising chain in a
dissipative cavity driven by an external transverse light field. When
driving
and dissipation
are in balance, the system can reach a nonequilibrium steady state which
undergoes a super-radiant phase transition as the driving strength increases.
Interestingly, the super-radiant field changes the effective bias of the Ising
chain in return and drives its own transition between the ferromagnetic and
paramagnetic phase. We study the rich physics in this system with sophisticated
behavior, and investigate important issues in its dynamics such as the stability
of the system and criticality of the phase transition.

\end{abstract}

\maketitle

A collection of two-level atoms in a cavity is a classical system for studying
atom-field interactions known as the Dicke model \cite{dicke1954coherence}. It hosts many
interesting physical effects including the unusual super-radiant phase
transition \cite{PhysRevA.8.2517} when the strength of the
atom-field interaction becomes so strong that it is comparable to the atomic
level splitting. Once the coupling strength exceeds a critical value, the ground
state of the system changes from vacuum to a state with nonzero macroscopic
photon occupation. There has been much interest in studying the super-radiant
phenomena in various contemporary context such as critical entanglement
\cite{PhysRevLett.92.073602}, finite-size scaling \cite{vidal2006finite}, and
quantum chaos \cite{PhysRevE.67.066203, PhysRevLett.90.044101}. Experimentally, it has been
observed in a system of laser-driven BEC coupled to an optical cavity
\cite{baumann2010dicke, brennecke2013real}. Though it has an origin in atomic and optical physics,
with the development of new technologies the Dicke model has found its place
in many other physical systems such as iron traps \cite{garraway2011dicke, genway2014generalized},
nitrogen-vacancy (NV) centers \cite{zou2014implementation},
superconducting qubits \cite{nataf2010no, garraway2011dicke}, and quantum dots
\cite{scheibner2007superradiance},
where the atoms and cavity can be replaced by qubits
and microwave circuits.

On a separate front, every quantum system is inevitably coupled to its
surrounding environment which is dissipative in nature. Though dissipation is
considered an obstacle in many quantum studies, it can also be actively
exploited to demonstrate interesting and nontrivial physics. For instance,
various nonequilibrium many-body phases and quantum phase transitions were
discovered \cite{PhysRevLett.105.015702, PhysRevLett.111.220408, lee2014dissipative, gelhausen2016quantum, tian2016cavity} under the balance of dissipation and driving. For the
Dicke model, the optical cavity is always leaky in reality and thus in principle
it is an open system. Because of this consideration, dynamical quantum phase
transitions in an open Dicke system has been investigated
\cite{PhysRevLett.104.130401, PhysRevA.84.043637, PhysRevLett.108.043003, bhaseen2012dynamics, PhysRevA.87.023831},
and it was found that such dynamical quantum phase transition can
exhibit different characteristics than the conventional Dicke model.

\begin{figure}
\includegraphics[width=1.0\linewidth]{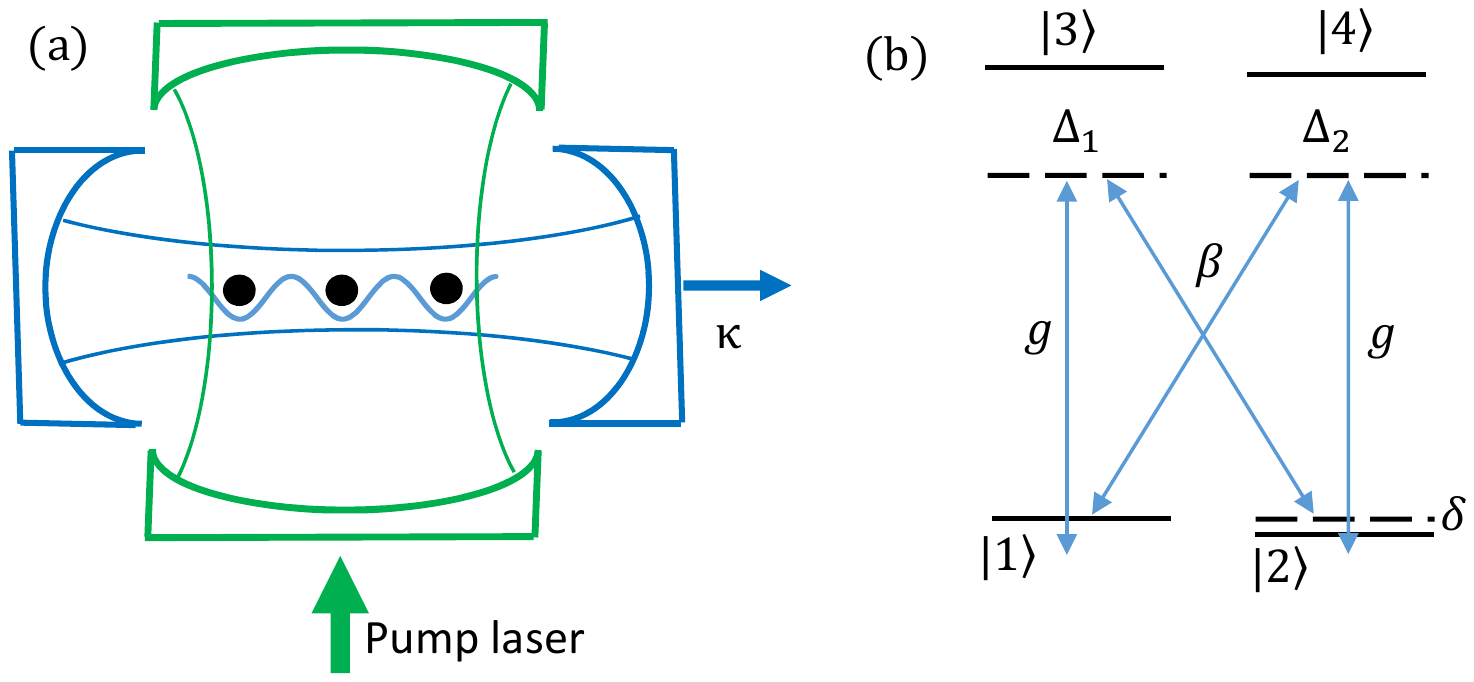}
\caption{\footnotesize{(a) The model system with a 1D chain of atoms with
nearest-neighbor interactions in a dissipative cavity and driven by an external
transverse field.
(b) Energy levels of the atoms and couplings between them by the cavity field
and external field. The detunings for the two Raman processes are approximately
equal. I.e., $\Delta_1\simeq\Delta_2\equiv\Delta_\text{e}$.}}
\label{Fig:model}
\end{figure}

Earlier work on the Dicke system has largely focused on weakly interacting atoms.
This is reasonable in an atomic BEC system since typically interactions between
neutral atoms in a cavity are weak. In many new physical systems such
as iron traps \cite{genway2014generalized} and solid-state qubits
\cite{nataf2010no, garraway2011dicke}, though, it is possible to
induce appreciable interactions between the two-level entities \cite{kim2010quantum, kim2011quantum, niskanen2007quantum} that play the
role of atoms in the original Dicke model. From a theoretical point of view,
introduction of strong interactions between the two-level entities is an interesting
addition to the system that can give rise to nontrivial new physics \cite{PhysRevLett.93.083001, gammelmark2011phase, zhang2014quantum, chen2016quantum}.
Not only
can it have an impact on the interaction between the two-level entities and
the cavity field, but it greatly enriches the physics of the system of the
two-level entities itself. Under such a consideration, in this work we study
the dynamical nonequilibrium quantum phase transition in a generalized Dicke
model with cavity leakage and Ising interactions between the two-level entities.
Specifically, as shown in Fig. \ref{Fig:model} (a), our system
consists of N identical atoms located inside a cavity and also driven by a
transverse external light field. Though we have used atoms to describe our
system, in reality they can be other entities such as ions \cite{genway2014generalized}
and solid-state qubits \cite{nataf2010no} depending on the specific physical system \cite{garraway2011dicke}.
It is assumed that the atoms
have two hyperfine ground states with an energy splitting of $\delta$ that are
coupled by the cavity mode and external driving field via two Raman processes \cite{PhysRevA.75.013804}
depicted in Fig. \ref{Fig:model} (b).
Further, atoms are arranged in a 1D chain structure with nearest-neighbor Ising interactions, which can be accomplished
using a quasi-1D optical lattice potential for atoms \cite{klinder2015observation, landig2016quantum, duan2003controlling} or simply by controlled
ion-trap potential for ion qubits \cite{kim2011quantum, genway2014generalized} and controlled
fabrication for solid-state qubits \cite{tian2010circuit, nataf2010no}.
Such a system
is described by the following Hamiltonian
of a coupled spin-field system with Ising interactions,
\begin{equation}
\begin{split}
H(t)=&(\omega_a-\frac{g^2}{\Delta_\text{e}}) a^{\dag}a+[\frac{g\beta(t)}{\Delta_\text{e}}a+\frac{g\beta^*(t)}{\Delta_\text{e}}a^{\dag}]\sum_i\sigma_i^x \\
&-\delta\sum_i\sigma_i^z-J\sum_i\sigma_i^y\sigma_{i+1}^y,
\end{split}
\end{equation}
where $a$ ($a^{\dag}$) is the annihilation (creation) operator for the cavity
field with frequency $\omega_a$, and $\sigma_i^x$, $\sigma_i^y$, and
$\sigma_i^z$ are Pauli matrices for an effective spin whose up and down states
are the two ground states in the atomic energy levels in Fig. \ref{Fig:model}
(b).
$\beta(t)=\beta_0\exp(i\omega_b t+i\varphi)$ describes the effect of the
driving field with frequency $\omega_b$ and phase $\varphi$, and $\Delta_e$ is
the detuning of the Raman processes in Fig. \ref{Fig:model} (b). $g$
characterizes the effective spin-field coupling strength, and $J$ is the
strength of Ising interactions between neighboring spins. In these parameters,
both the spin's energy splitting $\delta$ and the spin-field coupling strength
$g$ are much smaller than the frequency of the driving field and cavity mode,
$\omega_a$, $\omega_b$ $\gg$ $\delta$, $g\beta_0/\Delta_\text{e}$, $J$.
However, the detuning $\omega_a-\omega_b$ can be comparable to the spin's energy
splitting $\delta$ and interaction strength $J$.

To study the nonequilibrium phase transition in our system,
we first shift to the rotating frame defined by the external driving field by
applying the operator $U(t)=\exp(-i\omega_b a^{\dag}at)$. After
the Rotating Wave Approximation (RWA), we obtain the following effective
many-body time-independent Hamiltonian,
\begin{equation}
\begin{split}
H'=&(\Delta-i\frac{\kappa}{2}) a^{\dag}a+\frac{g_0}{\sqrt{N}}(a^{\dag}e^{i\varphi}+ae^{-i\varphi})\sum_i\sigma_i^x\\
&-\delta\sum_i\sigma_i^z-J\sum_i\sigma_i^y\sigma_{i+1}^y,
\end{split}
\label{eq:H'}
\end{equation}
where $\Delta=\omega_a-\frac{g^2}{\Delta_\text{e}}-\omega_b$ is the mismatch between the cavity and
external driving fields, $\kappa$ is the photon loss rate,
and $g_0=\sqrt{N}g\beta_0/\Delta_\text{e}$ \cite{PhysRevE.67.066203}. The
Hamiltonian in Eq. (\ref{eq:H'}), which contains both Ising interactions
between the spins and photon leakage out of the cavity, is an interesting and
useful model for studying dynamic phase transitions that has not been studied
so far.

In Eq. (\ref{eq:H'}), the second term dictates the interaction between the spin
chain and the cavity field assisted by the driving of the external field.
Because of the leakage of photons out of the cavity, the system evolves
irreversibly into a steady state until the effect of driving and dissipation
reaches a dynamical equilibrium. Obviously, this steady state has a decisive
impact on the physics of both the cavity field and Ising spin chain. To
determine the steady state, we expand the field operator $a$ and the spin-chain
operator $\sum_i \sigma_i^x$ around their mean field values $\phi_s$ and $S_x$
in the thermodynamic limit. This allows us to approximate the spin-field
interaction term as
\begin{equation}
\begin{split}
\frac{(a^{\dag}e^{i\varphi}+ae^{-i\varphi})}{\sqrt{N}}\sum_i\sigma_i^x\approx&\sqrt{N}(a^{\dag}e^{i\varphi}+ae^{-i\varphi})S_x\\
              &+2\phi_s\sum_i\sigma_i^x-2\phi_sS_xN,
\end{split}
\end{equation}
where $S_x=\frac{1}{N}\langle{\sum_i\sigma_i^x}\rangle_{ss}$ and
$\phi_s=\frac{1}{2\sqrt{N}}\langle{a^{\dag}e^{i\varphi}+ae^{-i\varphi}}
\rangle_{ss}$ are average values evaluated under the steady state.
The effective Hamiltonian can then be written as
\begin{equation}\label{Eq:meanfieldH}
\begin{split}
H_{\text{eff}}=&(\Delta-i\frac{\kappa}{2})
a^{\dag}a+g_0\sqrt{N}(a^{\dag}e^{i\varphi}+ae^{-i\varphi})S_x \\
&-2 g_0\phi_sNS_x
-\sum_i\Big(\delta\sigma_i^z-2g_0\phi_s\sigma_i^x+J\sigma_i^y\sigma_{i+1}
^y\Big).
\end{split}
\end{equation}

We can now derive the equations of motion for the field operators from
$H_{\text{eff}}$,
\begin{equation}\label{Eq:dissipation}
\begin{split}
\dot{a}=-i\Delta a-ig_0\sqrt{N}S_xe^{-i\varphi}-\frac{\kappa}{2} a\\
\dot{a}^{\dag}=i\Delta a^{\dag}+ig_0\sqrt{N}S_xe^{i\varphi}-\frac{\kappa}{2} a^{\dag},
\end{split}
\end{equation}
In the steady state
($\frac{d}{dt}a=0$), the mean photon field in cavity is
\begin{equation}\label{Eq:phis}
\phi_s=-\frac{\Delta g_0S_x}{\Delta^2+\kappa^2/4},
\end{equation}
which is related to the ground state average $S_x$ of the spin part of the
Hamiltonian \cite{zhang2014quantum}. Eq. (\ref{Eq:phis}) reveals a critical relation between the field
and spin-chain behavior resulting from the dynamic equilibrium. Specifically,
macroscopic occupation of cavity photons and macroscopic polarization of the
spins in the $x$ direction occur at the same time. Thus, a super-radiant
phase transition, if it occurs, may be accompanied by a phase transition in
the spin chain between the ferromagnetic and paramagnetic phase. To see how it
dictates the system characteristics, we notice that the mean field $\phi_s$ in
turn has a direct impact on the spin part of the Hamiltonian
\begin{equation}
H_\text{Ising}=-\sum_i\delta\sigma_i^z-2g_0\phi_s\sigma_i^x+J\sigma_i^y\sigma_{
i+1}^y
\label{Eq:Ising}
\end{equation}
by modulating the transverse field of the Ising model. The transverse field
Ising Hamiltonian in Eq. (\ref{Eq:Ising}), which has a ferromagnetic phase for a
strong Ising interaction $J>B_\bot=\sqrt{\delta^2 + 4g_0^2\phi_s^2}$, and a
paramagnetic phase for a strong transverse field $B_\bot>J$,
can be solved exactly by the Jordan-Wigner transformation
\cite{PhysRevLett.95.245701}. Its many-body ground state wave function, $\vert
\Psi_\text{g}(\phi_s)\rangle$, is dependent on the transverse field $B_\bot$ of the
Ising model and hence on $\phi_s$. From $\vert \Psi_\text{g}(\phi_s)\rangle$, we can
calculate the macroscopic polarization in the $x$ direction,
\begin{equation}
 S_x(\phi_s) = \frac{1}{N} \langle \Psi_\text{g}(\phi_s)\vert \sum_i \sigma_i^x \vert
\Psi_\text{g}(\phi_s) \rangle,
\label{Eq:Sx_phi}
\end{equation}
which is also dependent on $\phi_s$. By solving Eqs. (\ref{Eq:phis}) and
(\ref{Eq:Sx_phi}), we can then self-consistently determine the values of
$\phi_s$ and $S_x$ in the dynamical equilibrium.

\begin{figure}
\includegraphics[width=1.0\linewidth]{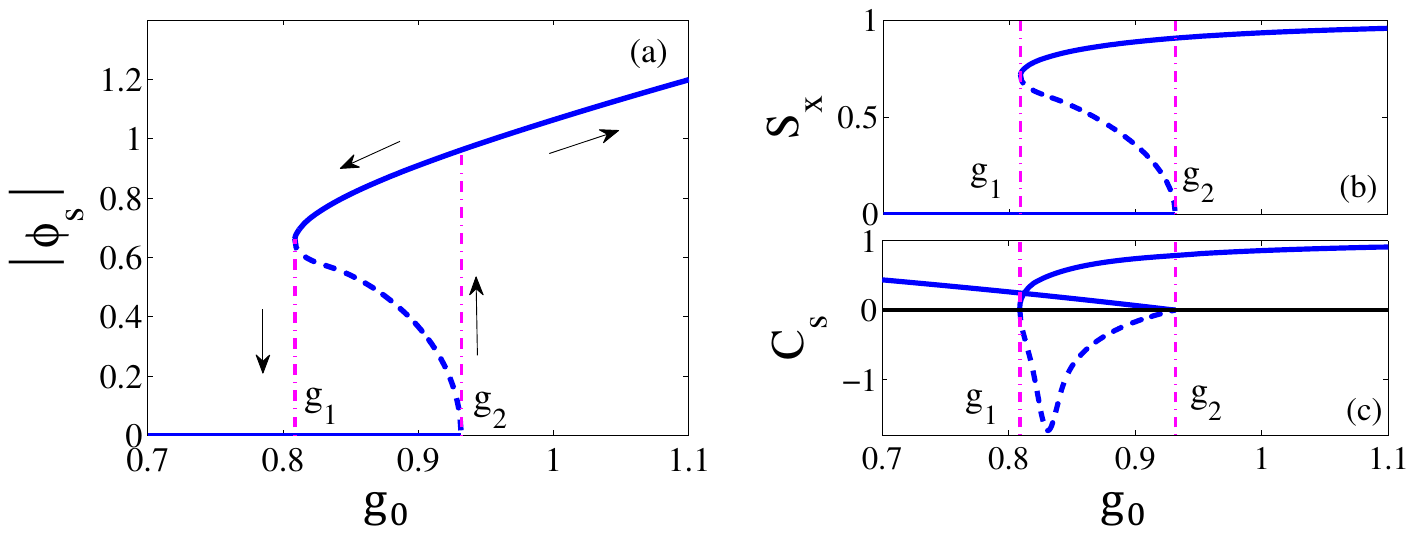}
\caption{\footnotesize{(a) Mean-field cavity photon numbers $\phi_s$ versus
driving strength $g_0$, with a hysteretic regime of bi-stable states. (b) Mean
polarization of the spin chain in the $x$ direction $S_x$ versus driving
strength $g_0$, with a stable paramagnetic phase (upper branch) and a stable
ferromagnetic phase (lower branch). (c) The stability coefficient $C_s$ versus
$g_0$. In all plots, stable branches are in solid lines and unstable branches are
in dashed lines. Other parameters are $\Delta=0.8$, $\delta=0.3$,
$\kappa=0.5$, $N=200$. We use $J$ as the energy unit.}}
\label{Fig:bistable}
\end{figure}

\begin{figure}
\includegraphics[width=1.0\linewidth]{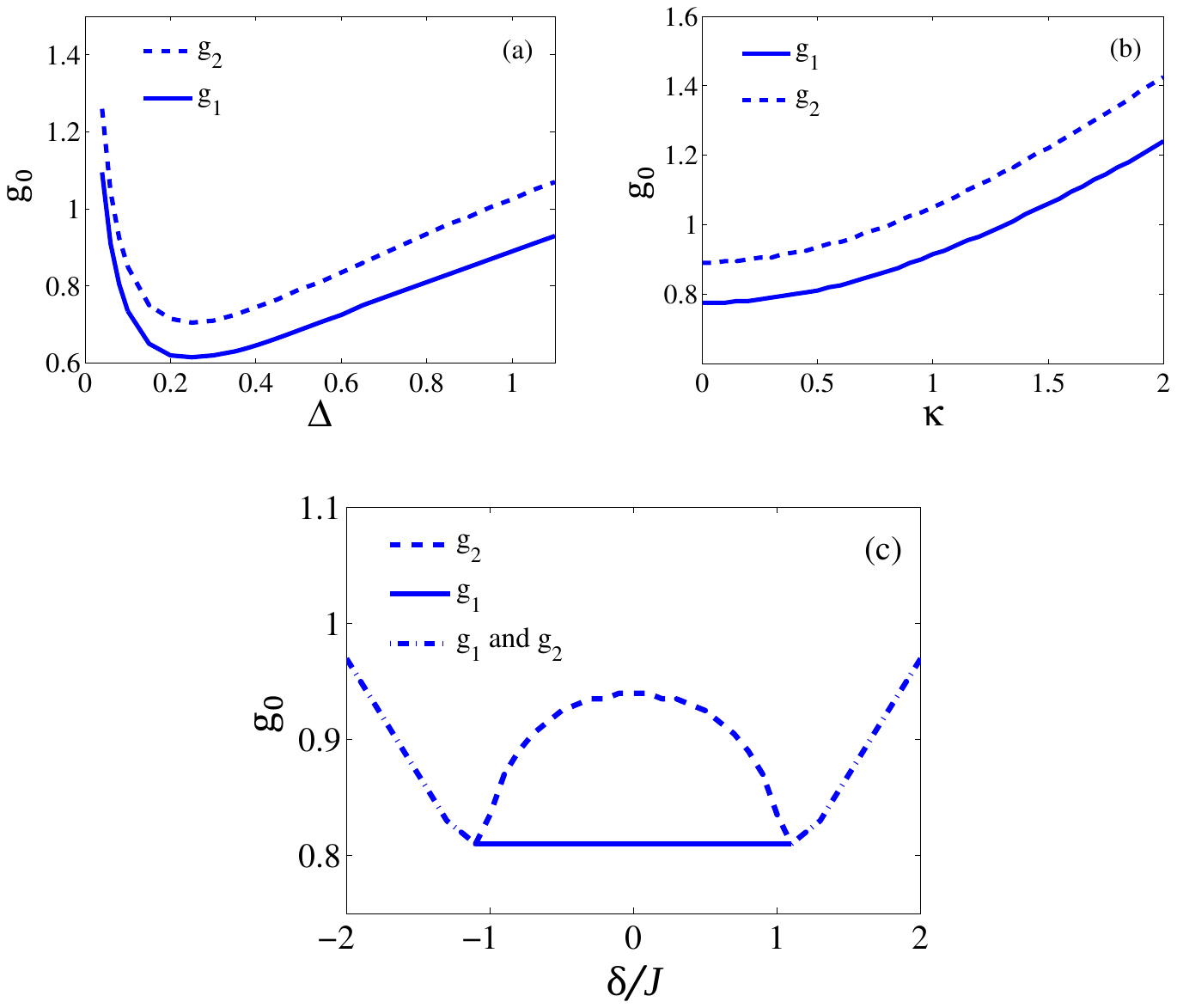}
\caption{\footnotesize{(a) Dependence of the upper and lower critical driving
strengths for the phase transitions on the detuning. (b) Dependence of the
critical driving strengths on the dissipation rate $\kappa$. (c) Dependence of
the critical driving strengths on the ratio between the energy splitting of the
spin and the Ising interaction strength, $\delta/J$.
The two critical driving strengths become equal
for $\delta/J$ greater than 1, indicating the disappearance of the hysteresis
in Fig. \ref{Fig:bistable}. Other parameters are the same as in Fig.
\ref{Fig:bistable} and $J$ is the energy unit.}}
\label{Fig:phase}
\end{figure}

In Figs. \ref{Fig:bistable} (a) and (b), we plot the solved values for $\phi_s$
and $S_x$ against the driving strength $g_0$
when the spin energy splitting $\delta$ is smaller than the Ising interaction
strength $J$. It is seen that,
when the coupling is weak, $\phi_s=0$, there is no macroscopic photon occupation
in the cavity. By Eq. (\ref{Eq:phis}), $S_x=0$, the spin-chain
does not exhibit a macroscopic polarization in the $x$ direction either. When
the coupling strength $g_0$ increases, $\phi_s$ becomes nonzero, indicating that
a super-radiant phase transition occurs. However, the solutions of $\phi_s$ and
$S_x$ are not single-valued in a range of driving strength $g_0$. The question
which solutions are physical can be resolved with a stability analysis.

To study the stability of the solutions for $\phi_s$ and $S_x$, we consider the
time evolution of deviations from their mean field value.
We split the field and spin operators into their steady state mean values and
small fluctuations
\begin{subequations}
\begin{align}
a(t)=&a_s+\delta a(t)\\
\frac{1}{N}\sum_i\sigma_i^x(t)=&S_x+\delta S_x(t),
\end{align}
\end{subequations}
with $a_s=\langle a\rangle_{ss}$.
Assuming a small fluctuation term for the field value, we have
$\phi=\phi_s+\delta
\phi$ with $\delta\phi=\frac{1}{2\sqrt{N}}(\delta
a^{\dag}e^{i\varphi}+\delta
ae^{-i\varphi})$,
and $\delta S_x(t)=\frac{\partial
S_x}{\partial \phi} \delta \phi$ \cite{tian2010circuit}.
Using the equation of motion
for the field operator in Eq. (\ref{Eq:dissipation}), we have
\begin{equation}
\dot{\delta \phi}=-\frac{\kappa}{2}C_s\delta \phi,
\end{equation}
where
\begin{equation}
 C_s=1+\frac{g_0}{\Delta^2+\kappa^2/4}\frac{\partial S_x}{\partial \phi}
\end{equation}
is the stability coefficient.
A solution for $\phi_s$ is stable if and only if $C_s>0$.
Therefore, from the plot of $C_s$ in Fig. \ref{Fig:bistable}(c), we
conclude that the branches in solid lines for the solutions of $\phi_s$ and
$S_x$ are stable, whereas the branches in dashed lines are unstable.

From this stability analysis, we can infer that there is a hysteresis in the
solutions for $\phi_s$ and $S_x$ with bi-stability as indicated in Figs.
\ref{Fig:bistable} (a) and (b). When the spin-field driving strength $g$ is
increased above a critical value $g_2$, the value of $\phi_s$ jumps from 0 to a
nonzero value, and the system makes a discontinuous transition to the
super-radiant phase. Likewise, the Ising spin chain experiences a discontinuous
phase transition at $g_2$ from the ferromagnetic phase to paramagnetic phase,
as shown in Fig. \ref{Fig:bistable} (b). Therefore, the phase transitions in
our dissipative system are first-order in nature. This is notably different
than the conventional transverse field Ising model, in
which the quantum phase transition is continuous \cite{PhysRevLett.95.245701}.
When the system is in the super-radiant phase and the driving strength is
decreased, it makes a transition to the vacuum state at a coupling strength
$g_1$ lower than $g_2$, giving rise to the hysteresis in Fig.
\ref{Fig:bistable}.

Studying the impact of the system's key parameters on the nature and
characteristics of phase transitions can help gain deep insight into our
dissipative system.
First, we calculate how the critical coupling strengths $g_1$ and $g_2$ for
the phase transitions are dependent on $\Delta$, the effective photon
frequency in the effective Hamiltonian in Eq. (\ref{eq:H'}). For a
non-dissipative Dicke system, the coupling strength required
for the super-radiant phase transition decreases monotonically with the photon
frequency \cite{zhang2014quantum}. As shown in Fig. \ref{Fig:phase} (a), the
behavior of our dissipative system is quite different. Not only do $g_1$ and
$g_2$ reach their minima at a finite value of $\Delta$, but they diverge when
$\Delta \rightarrow 0$. This can be understood from Eq. (\ref{Eq:phis}), which
suggests that the super-radiant phase transition cannot occur for $\Delta=0$
since $\phi_s=0$. As for the minimum values of $g_1$ and $g_2$, we can prove
that they occur at
$\Delta = \kappa/2$.
According to Eq. (\ref{Eq:phis}), the values of $g_1(\Delta=\kappa/2)$ and
$g_2(\Delta=\kappa/2)$ are determined by solving
\begin{equation}\label{eq:steady1}
\phi_s=-\frac{g_0S_x}{\kappa/2},
\end{equation}
with $S_x$ determined by the ground state of
\begin{equation}\label{eq:steady2}
H_\text{Ising}=-\sum_i\delta\sigma_i^z-2g_0\phi_s\sigma_i^x+J\sigma_i^y\sigma_{i+1}^y.
\end{equation}
For a value of $\Delta$ slightly different, $\Delta=\kappa/2\pm\epsilon$
($\epsilon \ll \kappa$), $g_1(\Delta=\kappa/2\pm\epsilon)$ and
$g_2(\Delta=\kappa/2\pm\epsilon)$
are determined by solving
\begin{equation}\label{eq:steady3}
\phi_s'=-\frac{g_0'S_x}{\kappa/2},
\end{equation}
with $S_x$ determined by the ground state of
\begin{equation}\label{eq:steady4}
H_\text{Ising}=-\sum_i\delta\sigma_i^z-2g_0'\phi_s'\sigma_i^x+J\sigma_i^y\sigma_{i+1}^y,
\end{equation}
where $g_0'$ and $\phi_s'$ are obtained by expansion of Eq. (\ref{Eq:phis}) to
second order of $\epsilon$ with the results
\begin{equation}
 g_0'=g_0\sqrt{1-2\epsilon^2/\kappa^2}
\end{equation}
and
\begin{equation}
 \phi_s'=\frac{\phi_s}{\sqrt{1-2\epsilon^2/\kappa^2}}.
\end{equation}
Notice that Eqs. (\ref{eq:steady3}) (\ref{eq:steady4}) are in the same form with
Eqs. (\ref{eq:steady1}) (\ref{eq:steady2}), indicating that their solutions are
related by a proper scaling. Specifically,
\begin{equation}
 g_1'(\Delta=\kappa/2\pm\epsilon)=g_1(\Delta=\kappa/2)
\end{equation}
with
\begin{equation}
 g_1'(\Delta=\kappa/2\pm\epsilon)\equiv
g_1(\Delta=\kappa/2\pm\epsilon)\sqrt{1-2\epsilon^2/\kappa^2}.
\end{equation}
Thus we have
\begin{equation}
g_1(\Delta=\kappa/2\pm\epsilon)=\frac{g_1(\Delta=\kappa/2)}{\sqrt{
1-2\epsilon^2/\kappa^2}} >g_1(\Delta=\kappa/2),
\end{equation}
and we can conclude that the values of $g_1$ and $g_2$ are the minimal at
$\Delta =\kappa/2$.

Then, we investigate the effect of dissipation rate $\kappa$ on the critical
coupling strengths $g_1$ and $g_2$ for the phase transitions. As shown in Fig.
\ref{Fig:phase} (b), $g_1$ and $g_2$ increase monotonically with the dissipation
rate $\kappa$, suggesting that the phase transitions occur at stronger
couplings when the dissipation is more severe. This is understandable, since a
higher rate of dissipation must be balanced by stronger driving to support the
super-radiant phase.
Finally, in Fig. \ref{Fig:phase} (c), we show how $g_1$ and $g_2$ change with
the relative magnitude of the spin energy splitting $\delta$ and Ising
interaction strength $J$. It is seen that, when $\delta$ is smaller than $J$,
$g_1$ and $g_2$ have different values, which suggests discontinuous first-order
phase transitions. However, when $\delta$ is greater than $J$, $g_1$ and $g_2$
are equal. This suggests that the hysteresis in Fig. \ref{Fig:bistable}
disappears. Closer examination of the
values of $\phi_s$ and $S_x$ reveals that the phase transitions have become
continuous. This behavior is similar to that in closed Dicke systems without
dissipation \cite{PhysRevLett.93.083001, gammelmark2011phase}, except that
$\delta/J$ needs to be slightly larger than 1 for the phase transitions to
become continuous due to the influence of the dissipation.

\begin{figure}
\includegraphics[width=1.0\linewidth]{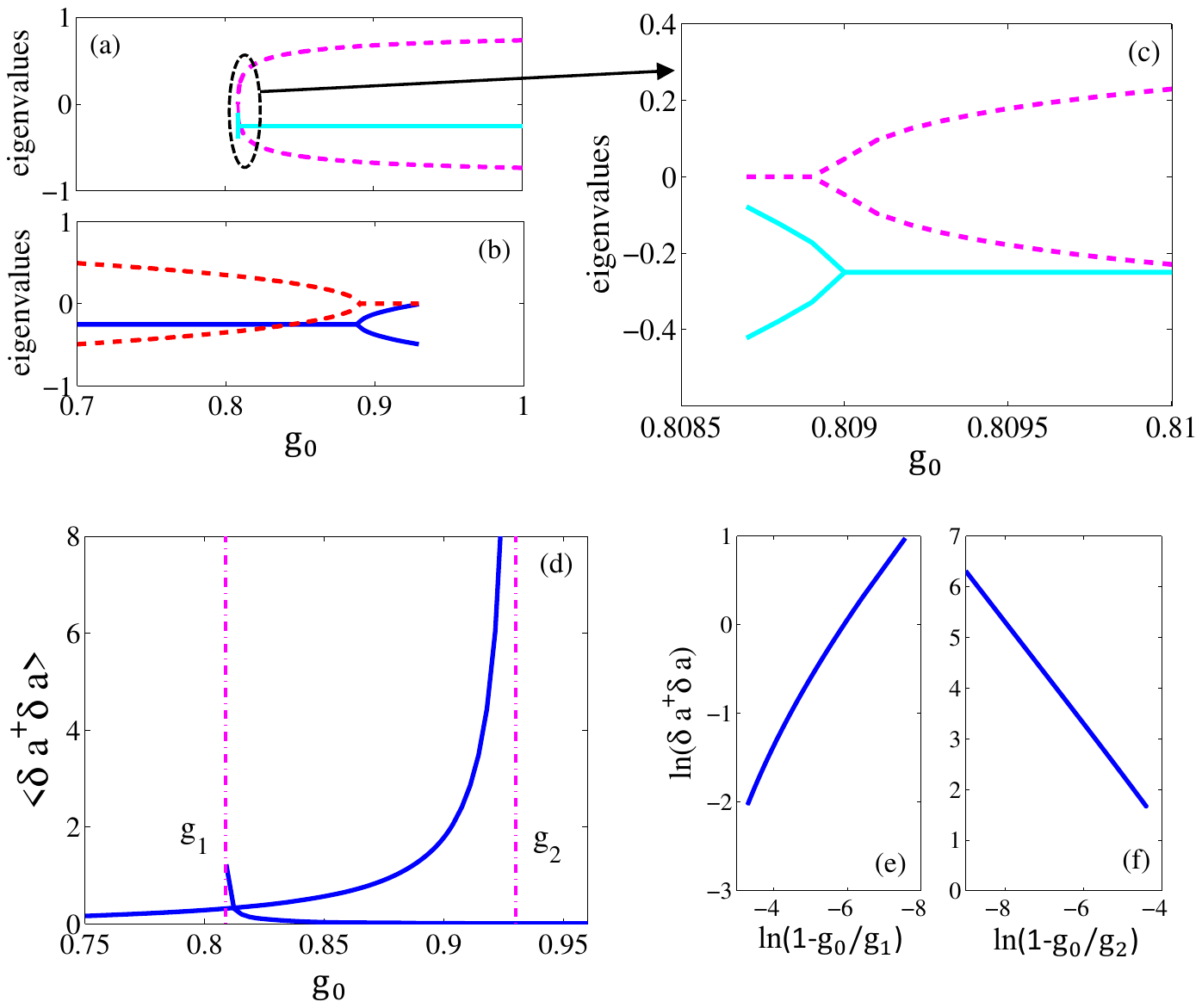}
\caption{\footnotesize{(a) Eigenvalues in Eq. (\ref{Eq:omega}) of the linear
stability matrix $\mathbf{M}$ for the stable paramagnetic phase.
The solid and dashed line are the real and image part respectively.
(b) Eigenvalues in Eq. (\ref{Eq:omega}) of the linear stability matrix
$\mathbf{M}$ for the stable ferromagnetic phase.
(c) Closer view of (a) around the critical driving strength $g_1$.
(d) The number of fluctuating photon number $\langle\delta a^{\dag}\delta a\rangle$ of the steady state when
the coupling strength approaches the critical points $g_1$ and $g_2$.
(e) (f) The fluctuating photon number $\langle\delta a^{\dag}\delta a\rangle$ on
a log scale. Parameters for the calculation are the same as in Fig.
\ref{Fig:bistable} and $J$ is the energy unit.
}}
\label{Fig:cp}
\end{figure}

One more characteristic of the system worth studying is fluctuations because
of the crucial role they play in phase transitions \cite{binder1987theory, plischke2006equilibrium}.
For simplicity, we focus on the fluctuation of the cavity field, and
assume that the Ising chain always stays in the ground state corresponding to
the cavity field. Under such an assumption, the fluctuation in the spin chain's
polarization in the $x$ direction, $S_x$, is $\delta S_x(t)=\frac{\partial
S_x}{\partial \phi} \delta \phi$ to first order in the cavity filed fluctuation
$\delta \phi$ \cite{tian2010circuit}.
Denoting the the field fluctuations with $\mathbf{V}=[\delta a, \delta
a^{\dag}]^T$, and using the equations of motion for the field operators, we have
\begin{equation}\label{eq:crit}
\frac{\partial}{\partial t}\mathbf{V}=\mathbf{M}\mathbf{V}+\widehat{\xi},
\end{equation}
where $\widehat{\xi}=(\xi,\xi^{\dag})^T$ is the vector of operators for the
quantum noise which has zero mean values and whose only non-vanishing
correlation is $\langle\xi(t)\xi^{\dag}(t')\rangle=\kappa\delta(t-t')$.
$\mathbf{M}$ is the linear stability matrix of the mean field solution
\begin{gather*}
\mathbf{M}=
\begin{bmatrix}
-i\Delta-i\frac{g_0}{2}\frac{\partial S_x}{\partial \phi_s}-\frac{\kappa}{2} & -i\frac{g_0}{2}\frac{\partial S_x}{\partial \phi_s}\\ i\frac{g_0}{2}\frac{\partial S_x}{\partial \phi_s} & i\Delta+i\frac{g_0}{2}\frac{\partial S_x}{\partial \phi_s}-\frac{\kappa}{2}
\end{bmatrix}
\end{gather*}

The matrix $\mathbf{M}$ is non-normal,
therefore it has different left and right eigenvectors $\mathbf{R}$, $\mathbf{L}$ \cite{PhysRevA.84.043637},
which form a biorthogonal system $\mathbf{L}\mathbf{R}=\mathbf{I}$.
$\mathbf{M}$ has
two eigenvalues,
\begin{subequations}
\begin{align}
\omega_{m1}=\frac{-\kappa-i\sqrt{4\Delta^2+4\Delta g_0 \partial S_x/\partial \phi}}{2}\\
\omega_{m2}=\frac{-\kappa+i\sqrt{4\Delta^2+4\Delta g_0 \partial S_x/\partial \phi}}{2}.
\end{align}
\label{Eq:omega}
\end{subequations}
which are plotted as a function of driving strength in Figs. \ref{Fig:cp}(a)
and (b). When the real part of the eigenvalues is negative, the
system is stable \cite{PhysRevA.84.043637}.
Following the method discussed in Ref. \cite{PhysRevA.84.043637}, we solve for
the number of fluctuating photon number which is
\begin{equation}
\langle\delta a^{\dag}\delta a\rangle=-\sum_{i,j}\frac{\kappa}{\omega_{mi}+\omega_{mj}}L_{i,1}L_{j,2}R_{2,i}R_{1,j}
\end{equation}
where $\omega_{mi}$, $\omega_{mj}$ are the eigenvalues of matrix $\mathbf{M}$.
As shown in Fig.
\ref{Fig:cp} (d), when the driving strength decreases and approaches the
critical point $g_1$ of the system, the photon number fluctuation in the steady state
becomes
significant, which signals that the photon field enters a normal phase
from a super-radiant phase.
Likewise, when the driving strength increases and approaches the critical
value $g_2$, the fluctuation grows quickly.
In Figs. \ref{Fig:cp}(e) and (f), we plot the photon fluctuations
on a log scale as a function of deviation of the driving strength from the
critical point $g_1$ and $g_2$. The critical exponent can be read from these
plots. Near $g_2$, it is $-1.0$, consistent with the super-radiant phase
transition in a conventional Dicke model without Ising interaction. Near
$g_1$, the exponent is about $-0.75$.

In summary, we have studied dynamic phase transitions in an open system
consisting of an Ising chain in a lossy cavity. When the external driving and
cavity leakage reach a dynamic equilibrium, the cavity can undergo a
super-radiant phase transition when the driving strength for the Ising chain
is strong enough. Interestingly, because of the mutual impact between
the Ising chain and cavity field, it is accompanied by a transition between
ferromagnetic and paramagnetic phase in the Ising chain.
Under certain conditions, the phase transitions are hysteretic
and discontinuous, and exhibit different characteristics than those in
conventional closed systems. By studying important issues of the system such as
stability and fluctuations, we gained valuable insights in the unique properties
of quantum phase transitions in driven systems in dynamic equilibrium.


\end{document}